# Synchronous locating and imaging behind scattering medium in a large depth based on deep learning

SHUO ZHU,[1,†] ENLAI GUO,[1,†] QIANYING CUI,[1] DONGLIANG ZHENG,[1] LIANFA BAI,[1,*] AND JING HAN,[1,*]

[1]Jiangsu Key Laboratory of Spectral Imaging and Intelligent Sense, Nanjing University of Science and Technology, Nanjing 210094, China
* mrblf@163.com, * eohj@njust.edu.cn

**Abstract**: Scattering medium brings great difficulties to locate and image planar objects especially when the object has a large depth. In this letter, a novel learning-based method is presented to locate and image the object hidden behind a thin scattering diffuser. A multi-task network, named DINet, is constructed to predict the depth and the image of the hidden object from the captured speckle patterns. The provided experiments verify that the proposed method enables to locate the object with a depth mean error less than 0.05 mm, and image the object with an average PSNR above 24 dB, in a large depth ranging from 350 mm to 1150 mm. The constructed DINet can obtain multiple physical information via a single speckle pattern, including both the depth and image. Comparing with the traditional methods, it paves the way to the practical applications requiring large imaging depth of field behind scattering media.

Recovering hidden object through scattering medium has the broad application prospects in many fields ranging from atmospheric optics to biomedical optics [1, 2]. However, scattering media disrupts original information of the observed objects, and brings challenges to the imaging and measurement [3]. Some methods have been put forward to solve the scattering imaging problem, such as single-pixel imaging [4, 5], wavefront shaping imaging [6-8] and optical memory effect (OME) imaging [9-11]. However, the single-pixel imaging methods need more acquisition times with different compression ratios on different scenes, the state-of-the-art devices cannot shape the complex-valued wavefront precisely via wavefront shaping methods and the OME methods have a strict limitation on the field of view. Instead of building a complex physical model, deep learning (DL) is efficient to solve complex mapping relationships, which can generate an optimized model driven by a large-scale dataset [12]. The applications in scattering objects is a classical problem with complex mapping principle. For imaging through scattering media, DL has been successfully demonstrated to reconstruct through ground glasses, multimode fibers and fat emulsion [13-17]. The DL methods have great capability in generalization [15] and can also expend at least 40 times of the OME range [18].

The speckle patterns restoration is the main focus of the above research works, rather than other physical information. The object ranging and positioning are essential to the atmosphere or biological applications. To date, some techniques can acquire the hidden object depth via WFT-based (Windowed Fourier Transforms) ME measurement in phase space [19], PSF manipulation [20], chromatic aberration compensation [21] and coherence gating [22]. In order to obtain the depth information, the traditional physical methods are realized indirectly through the mapping relationship between the speckle pattern and the planer object position. Thus, these methods listed above cannot obtain depth and reconstructed image synchronously. The chromatic aberration compensation and coherence gating methods have good performance in traditional methods, which need additional optical reference arm to provide depth-related reference information. The reference arm configuration will make the experimental arrangement and adjustment more complicated. Besides, these depth detection methods are difficult to build complete physical models for obtaining absolute depth information. In which, the reported maximum depth is 90.5 mm on one side via chromatic aberration compensation. Further expansion of the depth detection capabilities is meaningful for expanding the practical applications. The critical requirements to acquire these physical information in these detection methods restrict their wide application in practice. However, the DL method on depth detection is limited with physical instruction and effective network structure. The depth information of hidden planar objects cannot be measured efficiently, with or without prior information, due to the scattering disruption and the model ability.

In this letter, for the first time, a novel DL framework (DINet) is proposed to realize simultaneous depth prediction and image reconstruction from a single speckle pattern, which is a dual-channel network providing different attributes. Unlike usual DL applications, where the information distribution or degradation degree is changed, requiring the DINet has great capability for data mining and multi-task collaboration. DINet can also simplify the hardware requirements and experimental process by removing the reference arm configuration. To the best of our knowledge, this is the first model that solves the problem of quantitative locating and synchronous imaging in field up to 1150 mm through scattering.

The end-to-end DINet is proposed to learn a statistical model relating to the speckle patterns generated in different positions. The practical systematic configuration, which is designed to collect the experimental data including the speckle patterns and the distance between object and diffuser, is drawn schematically in Fig. 1(a). As for Fig. 1(b), it is the description of locating distance with optical path unfold. The content structure of DINet is shown in Fig. 1(c). A single speckle pattern goes through the dual-channel network, which produces a depth value by the locating-channel network and a 256*256 reconstruction with the imaging-channel network.

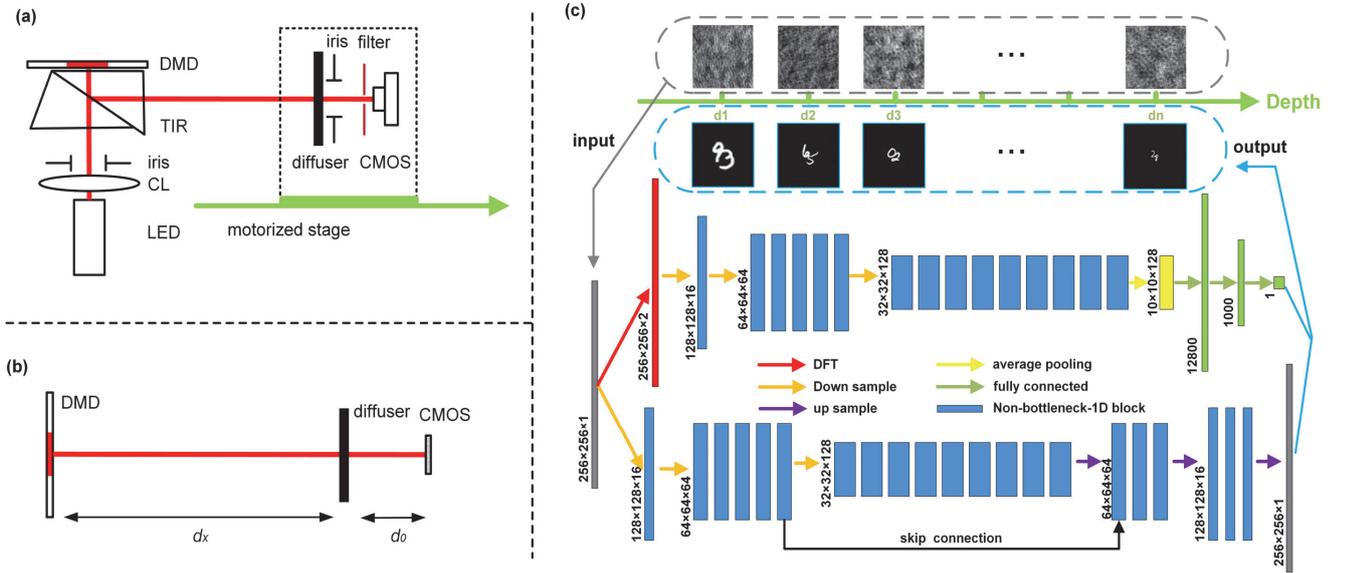

Fig.1. Systematic configuration. (a) Experimental arrangement. CL, collimating lens; TIR, total internal reflection prism; DMD, digital micro-mirror device. (b) Description of imaging distance with optical path unfold. (c) Schematic diagram of DINet Architecture.

The key structure of the multi-task network is based on Efficient Residual Factorized (ERF) layers [23], which is a residual architecture with reducing the computational costs and remaining the remarkable accuracy. The locating-channel consists of Discrete Fourier Transform (DFT), encoder part and fully connected layers to extract the location features and regress the depth values. The imaging-channel follows the encoder-decoder architecture with modification of the long-range skip connection to improve reconstruction quality.

To train the DINet, the mean absolute error (MAE) and the mean squared error (MSE) is used as loss function for evaluating the training model. The MAE for the depth-channel and the MSE for the imaging-channel are calculated as:

$$Loss = Loss_D + Loss_I = \frac{1}{N}\sum_i^N \|D_i - D_{gt}\| + \frac{1}{N}\sum_i^N \|I_i - I_{gt}\|^2 \quad , \qquad (1)$$

where $D_i$ and $D_{gt}$ are the predicted values and true values, $I_i$ and $I_{gt}$ are the reconstructed images and ground truths, respectively; $i$ is the index number of the training dataset, and $N$ is the mini-batch size. Two sub-networks can be trained synchronously with multi-task loss in the DINet framework. If only a particular function needed, the sub-channel network can also be trained and work independently.

Phase-space measurement of scattering provides the features for depth information calibration [19]. Phase-space optics contains the spatial and spatial frequency information, which allows the visualization of space position information [24]. The Wigner distribution function (WDF) can be used to describe the phase space features, defined as

$$f(r,k) = \int \left\langle \psi^*(r+\xi/2)\psi(r-\xi/2) \right\rangle e^{ik\xi} d\xi \quad , \tag{2}$$

where $r = (x, y)$ and $k = (k_x, k_y)$ are the two-dimensional spatial and spatial frequency vectors, and $\psi(r)$ is the wave spread function [25, 26]. Due to the visible phenomenon that depth variation would make the Phase-space images changing regularly, the fitted slope of speckle patterns, $A$, is related to variable depth linearly as $A \propto 1/d_x$. Thus, the DFT process of speckle patterns can help locating-channel network to extract the distance features for depth prediction. The DINet can regress the depth value via effective data mining and powerful fitting capability, which can optimize the complex mapping relationships as:

$$d_x = \mathbb{F}(S, d_0, \lambda) \quad , \tag{3}$$

where $\mathbb{F}$ is mapping function generated by DINet, $S$ is the speckle pattern of hidden object, $d_0$ is the distance between CMOS and scattering media, and $\lambda$ is the wavelength of light source.

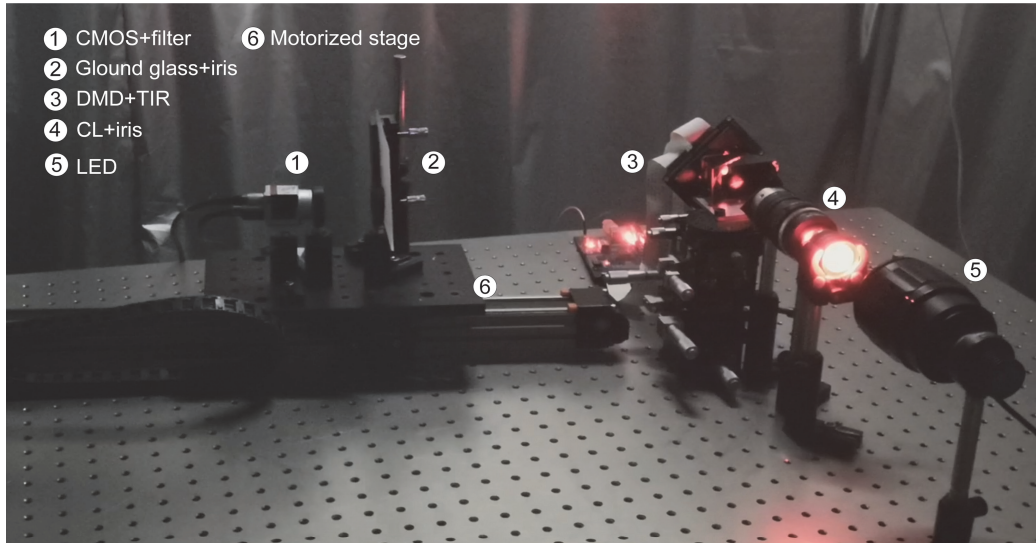

Fig.2. Schematic of Speckle collecting setups.

Finally, the proposed DINet is tested on real optical datasets and the datasets acquisition system is displayed in the Fig. 2. A digital micro-mirror device (DMD) (pixel count: 1024*768, pixel pitch: 13.68 μm) is employed to display handwritten digits selected as object images from the MINIST database on its surface. A TIR prism is employed to fold light path for capturing the patterns conveniently. A ground glass diffuser (Thorlabs, DG100X100-220-N-BK7) is selected as the thin scattering diffuser. The LED (Thorlabs, M625L4) combined with a filter (Thorlabs, FL632.8-1, central wavelength: 632.8±0.2 nm) is designed as the narrow band partially coherent light source for the experimental arrangement. The speckle patterns corresponding to different positions can be obtained by moving the motorized stage. As shown in Fig. 1(b), the optical path can be unfolded from TIR prism between the DMD and the CMOS (Balser, acA1920-155um). The value of $d_0$ is 80 mm which is the distance between ground glass and CMOS camera working surface. The devices in the dashed box are fixed relatively to ensure the $d_0$ is a constant value. Thus, the experimental setup within the dashed box are moved within the working stroke by motorized stage to get the variable depth of the hidden planar objects.

For training the DINet, 1100 speckle patterns are recorded at each position and selecting 1000 speckle patterns as training data, 50 speckle patterns as validation data, and 50 speckle patterns as testing data. The training set is processed with a mini-batch size of 32. Each model is trained with 400 epochs by Adam optimizer for up to 8 hours. The learning rate starts with $5*10^{-4}$ in the first 200 epochs and $5*10^{-5}$ for the final 200 epochs. The DINet is performed using the PyTorch 1.3.1 Python library on a single NVIDIA GeForce Titan RTX graphics unit.

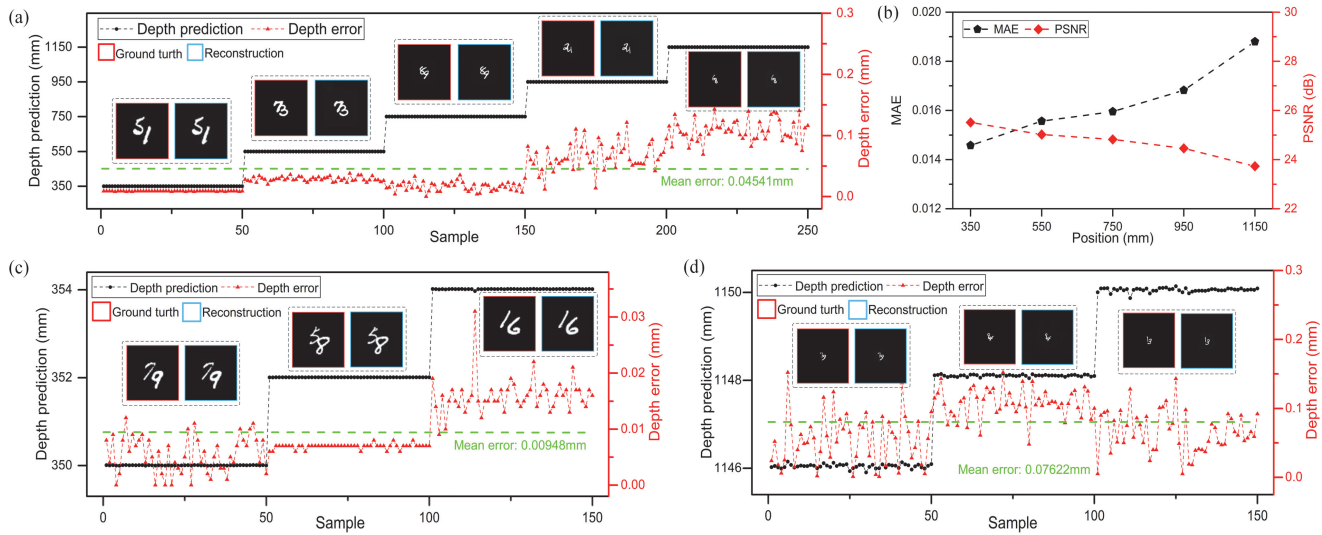

Fig.3. Quantitative evaluation for DINet. (a) The depth predictions and errors in 5 positions. An object ground truth and corresponding reconstruction are shown in each sample range. (b) The average MAE and PSNR are calculated for imaging evaluation each position. (c) and (d) are the resolution capability of the DINet in different ranges.

To quantitatively evaluate the locating and imaging performance of the DINet, the MAE and peak signal-to-noise ratio (PSNR) is employed to measure the locating accuracy and imaging quality. The multi-task results of untrained samples in the whole working stroke are shown in the Fig. 3(a), which include the locating testing results and imaging examples with corresponding ground truth. The distance of object relative movement from 350 mm to 1150 mm with stride of 200 mm. The abscissa is testing sample sequence and each position has 50 untrained samples. Obviously, the DINet can regress the depth of different distribution and restore the hidden object. The locating accuracy and error distribution are represented clearly via comparing with mean depth error, which is 0.04541mm indicated with green dash line. From the depth error distribution, the locating accuracy is reducing with distance increasing. Meanwhile, the result of the imaging-channel evaluation via MAE and PSNR is drawn in Fig. 3(b). The imaging quality is reducing slightly with distance increasing due to too small targets, and the average PSNR of reconstruction is calculated up to 24.7 dB in the whole working stroke.

The DINet has subdivision capability to locating and imaging from large stride to slight range. Both the ends of working stroke are selected to test DINet by changing the stride to 2 mm. From Fig. 3(c) and (d), the DINet can keep high resolution in the beginning stroke (350-354 mm) on the multi-task. Corresponding to the beginning stroke, the DINet can also complete the multi-task during the ending stroke (1146-1150 mm) of the subdivided working stroke. But, the refining capability of the ending stroke is lower than beginning in case of the same stride, with a mean error declining from 0.00948 mm to 0.07622 mm. It is consistent with the above analysis of the error distribution. To conclude, the DINet has good performance in locating and imaging from Fig. 3(a) and (b). However, with the distance increasing, the quantity of collected speckle patterns is limited by CMOS sensitivity. Besides, more system noise will be introduced, such as stray light and stage collimation error. All of above factors will result in performance reducing for locating accuracy and imaging quality.

When the depth of the target is not placed in the exact position as the training set, the DINet still has the depth generalization capability to recover targets and their depth synchronously. The DINet has different generalization capability with the distribution of position and step size. As shown in the Table 1 and the Fig. 4, the generalization of missing distance in the middle is better than the marginal and the smaller sampling steps is better than the large relatively. The depth generalization capability decreases with distance. It is difficult to obtain generalization depth information and reconstruction with too large step size. The results of depth generalization capability can also be explained from the aspect of data acquisition, which corresponds to the distribution of uniform sampling and the effective interval of sampling points.

Table 1. Results of depth generalization

| Training depth (mm) | Generalize depth (mm) | Mean error (mm) |
|---|---|---|
| 350, 352, 356, 358 | 354 | 0.2143 |
| 352, 354, 356, 358 | 350 | 0.6612 |
| 350, 550, 950, 1150 | 750 | 48.3811 |

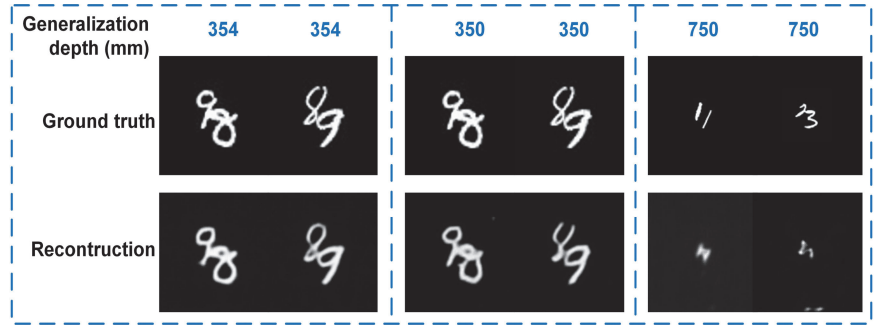

Fig.4. Reconstruction results in generalization depth.

The DINet can obtain relatively accurate generalization via the subdivision ability of dense measurement and extending the coverage of distance. The multi-modal measurement technology through a single speckle pattern based on the DINet has potential in practical applications, e.g., removing the effect of inclement weather conditions to produce a photo and a depth map in fog. The capability of depth generalization makes the DINet pave the way to a practical application.

In addition, the locating capability of DINet has nothing to do with the sizes of the images in the dataset. In geometrical optics, the transverse magnification of an optical system is given by the ratio of the image size to the object size, as well as the image distance to the object distance, $m = h'/h = v/u$, where $m$ is the magnification, the object size and the image size are $h$ and $h'$, and the object distance and the image distance are $u$ and $v$, respectively. If we set a constant target on the DMD, the size of ground truth captured without scattering medium is proportionally decreasing with the distance between the camera and the DMD. Meanwhile, the size of speckle autocorrelation behind scattering medium is also proportionally decreasing with the distance between the diffuser and the DMD [27]. However, the distance $d_x$ is the only variable related to the recording position of the speckle patterns via the DINet. In order to obtain the same imaging size, the 80 pixel objects and the 102 pixel objects are selected in the 350 mm position and the 450 mm position, respectively. The depth regression and image reconstruction are shown in Fig. 5. It is proved that the locating capability of DINet is only relevant to the location of object and has nothing with the size of the object.

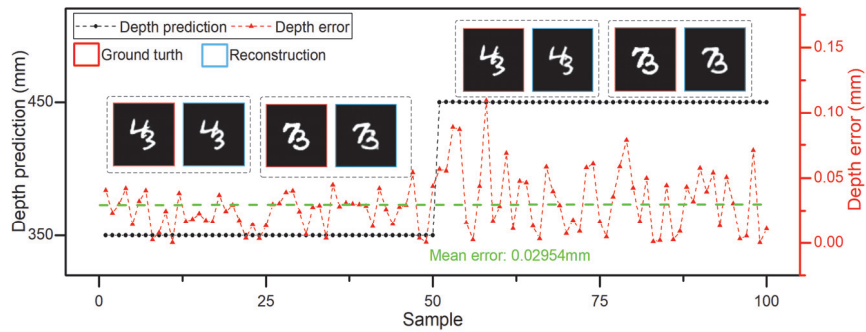

Fig.5. Locating and imaging results with the different object size and the same imaging size.

In this Letter, a multi-task DINet is designed for simultaneous prediction of depth information and reconstruction of object image from a single speckle pattern. This new method obtains reliable depth and imaging results, with an average error less than 0.05 mm in depth prediction and an average PSNR above 24 dB in object reconstruction, up to 1150 mm. This method, which is not limited to the discussed two tasks, does not need a complex experimental setup and can be applied to measure other physical information, such as locating in plane coordinates and hidden object classification. The proposed DINet technique opens up the way to multiple physical information measurement in practical application in case of scattering. But, the DINet can only regress the plane depth value in case of scattering. In the future, for pixel locating or three-dimensional measurement, a prior physical model will be helpful to guide the design of DL method.

†These authors contributed equally to this work.

**Acknowledgment**: This experimental arrangement work was supported by Yan Sun, Yingjie Shi and Jie Gu. The anthors thank Xiaoyu Chen, Bei Sun and Zhao Zhang for DL methods discussions.

## References

1. Ntziachristos, Vasilis. "Going deeper than microscopy: the optical imaging frontier in biology." Nature methods 7.8, 603 (2010).
2. Gibson, A. P., J. C. Hebden, and Simon R. Arridge. "Recent advances in diffuse optical imaging." Physics in Medicine & Biology 50.4, R1 (2005).
3. J. W. Goodman, Speckle Phenomena in Optics: Theory and Applications (Roberts & Company, 2007).
4. E. Tajahuerce, V. Durán, P. Clemente, E. Irles, F. Soldevila, P. Andrés, and J. Lancis, "Image transmission through dynamic scattering media by single-pixel photodetection," Opt. Express 22, 16945–16955 (2014).
5. Y.-K. Xu, W.-T. Liu, E.-F. Zhang, Q. Li, H.-Y. Dai, and P.-X. Chen, "Is ghost imaging intrinsically more powerful against scattering?" Opt. Express 23, 32993–33000 (2015).
6. I. M. Vellekoop and A. Mosk, "Focusing coherent light through opaque strongly scattering media," Opt. letters 32, 2309–2311 (2007).
7. S. Popoff, G. Lerosey, R. Carminati, M. Fink, A. Boccara, and S. Gigan, "Measuring the transmission matrix in optics: an approach to the study and control of light propagation in disordered media," Phys. Review letters 104, 100601 (2010).
8. A. Drémeau, A. Liutkus, D. Martina, O. Katz, C. Schülke, F. Krzakala, S. Gigan, and L. Daudet, "Reference-less measurement of the transmission matrix of a highly scattering material using a dmd and phase retrieval techniques," Opt. express 23, 11898–11911 (2015).
9. J. Bertolotti, E. G. Van Putten, C. Blum, A. Lagendijk, W. L. Vos, and A. P. Mosk, "Non-invasive imaging through opaque scattering layers," Nature 491, 232–234 (2012).
10. O. Katz, P. Heidmann, M. Fink, and S. Gigan, "Non-invasive singleshot imaging through scattering layers and around corners via speckle correlations," Nat. photonics 8, 784 (2014).
11. A. Porat, E. R. Andresen, H. Rigneault, D. Oron, S. Gigan, and O. Katz, "Widefield lensless imaging through a fiber bundle via speckle correlations," Opt. express 24, 16835–16855 (2016).
12. Y. LeCun, Y. Bengio, and G. Hinton, "Deep learning," nature 521, 436–444 (2015).
13. S. Li, M. Deng, J. Lee, A. Sinha, and G. Barbastathis, "Imaging through glass diffusers using densely connected convolutional networks," Optica 5, 803–813 (2018).
14. M. Lyu, H. Wang, G. Li, S. Zheng, and G. Situ, "Learning-based lensless imaging through optically thick scattering media," Adv. Photonics 1, 036002 (2019).
15. Y. Li, Y. Xue, and L. Tian, "Deep speckle correlation: a deep learning approach toward scalable imaging through scattering media," Optica 5, 1181–1190 (2018).
16. N. Borhani, E. Kakkava, C. Moser, and D. Psaltis, "Learning to see through multimode fibers," Optica 5, 960–966 (2018).
17. Y. Sun, J. Shi, L. Sun, J. Fan, and G. Zeng, "Image reconstruction through dynamic scattering media based on deep learning," Opt. express 27, 16032–16046 (2019).
18. E. Guo, S. Zhu, Y. Sun, L. Bai, C. Zuo, and J. Han, "Learning-based method to reconstruct complex targets through scattering medium beyond the memory effect," Opt. Express 28, 2433–2446 (2020).
19. K. T. Takasaki and J. W. Fleischer, "Phase-space measurement for depth-resolved memory-effect imaging," Opt. express 22, 31426–31433 (2014).
20. X. Xie, H. Zhuang, H. He, X. Xu, H. Liang, Y. Liu, and J. Zhou, "Extended depth-resolved imaging through a thin scattering medium with psf manipulation," Sci. reports 8, 1–8 (2018).

21. J. Xie, X. Xie, Y. Gao, X. Xu, Y. Liu, and X. Yu, "Depth detection capability and ultra-large depth of field in imaging through a thin scattering layer," J. Opt. 21, 085606 (2019).
22. O. Salhov, G. Weinberg, and O. Katz, "Depth-resolved specklecorrelations imaging through scattering layers via coherence gating," Opt. letters 43, 5528–5531 (2018).
23. E. Romera, J. M. Alvarez, L. M. Bergasa, and R. Arroyo, "Erfnet: Efficient residual factorized convnet for real-time semantic segmentation," IEEE Transactions on Intell. Transp. Syst. 19, 263–272 (2017).
24. L. Waller, G. Situ, and J. W. Fleischer, "Phase-space measurement and coherence synthesis of optical beams," Nat. Photonics 6, 474 (2012).
25. M. A. Alonso, "Wigner functions in optics: describing beams as ray bundles and pulses as particle ensembles," Adv. Opt. Photonics 3, 272–365 (2011).
26. M. J. Bastiaans, "Applications of the wigner distribution function to partially coherent light beams," in Selected Papers from International Conference on Optics and Optoelectronics' 98, vol. 3729 (International Society for Optics and Photonics, 1999), pp. 114–128.
27. C. Guo, J. Liu, T. Wu, L. Zhu, and X. Shao, "Tracking moving targets behind a scattering medium via speckle correlation," Applied optics, vol. 57, no. 4, pp. 905–913, 2018.